# EFFECTS OF THE PLANAR GALACTIC TIDES AND STELLAR MASS ON COMET CLOUD DYNAMICS

M. MASI[a], L. SECCO[a] AND G. GONZALEZ[b]

[a]*Dipartimento di Astronomia, Vic. dell'Osservatorio 2, 35122 Padova, Italy*

[b]*Grove City College, Rockwell Hall, 100 Campus Drive, Grove City, PA 16127 USA*

Editorial correspondence to:

Masi Marco

*Dipartimento di Astronomia, Vic. dell'Osservatorio 2, 35122 Padova, Italy*

E-mail address: marco.masi@gmail.com

**Abstract.** We report the first results of a research program to explore the sensitivity of the orbits of Oort cloud comets to changes in the strength of the Galactic tides in the plane of the disk and also to changes in the mass of the host star. We performed 2D simulations that confirm that the effects of the tides on comet orbits are sensitive to a star's distance from the Galactic center. A comet cloud closer to the Galactic center than the Sun will have comet perihelia reduced to the region of the inner planets more effectively by the planar tides alone. Similar results are found for a star of smaller mass. We also show how this phenomenon of comet injection persists for a set of alternative Galactic potential models. These preliminary results suggest a fruitful line of research, one that aims to generalize the study of comet cloud dynamics to systems different from the Solar System. In particular, it will allow us to study the roles played by comet clouds in defining the boundaries of the Galactic Habitable Zone.

# 1. Introduction

Were special conditions required for the formation and evolution of our habitable Solar System? Which particular properties of the Solar System are important for the existence of life on Earth? These are questions astrobiologists are beginning to address through observations and simulations. Research on extrasolar planets, for example, has revealed that their properties vary over a much broader range than had been known or anticipated from study of the properties of the planets in our Solar System. The Galactic context is an important factor in determining the properties of a planetary system [1,2]. Obviously, the Sun's location and orbit in the Galaxy are different from any other star. But, which aspects of its Galactic context are relevant to its habitability? More specifically, how do the dynamical properties of an Oort cloud depend on Galactic location? It is this question we will begin to address in the present work.

## 1.1 COMETS AND THE GALACTIC HABITABLE ZONE

Several Galactic-scale factors can influence the habitability of a planet [2]. They include factors relevant to the formation of planets, such as the radial disk metallicity gradient, and events that can threaten life on a planet, such as nearby supernovae, gamma ray bursts and comet impacts.

Catastrophic events have profoundly influenced the history of life on Earth, as is evidenced, for example, in the paleontological record of mass extinctions [3]. Impacts of asteroids and comets on Earth are plausible causes of these events. But, some researchers think that comets might also have played an important role in delivering the prebiotic organic compounds necessary for the origin of life on Earth (for instance, the Stardust spacecraft recently captured organics from comet 81P/Wild 2 [4]). Interplanetary dust particles from asteroids and comets indeed contain organics that can survive atmospheric entry [5]. Comets have possibly played a role in delivering the water in Earth's oceans [6]. This seemed at first implausible since the deuterium/hydrogen ratios (D/H) in the P/Halley, Hyakutake and Hale-Bopp comets [7-9] are about twice that of terrestrial water, which has been generally believed to originate from carbonaceous chondrites (having similar D/H ratios). However, [10] point out that taking present isotopic abundances as indicators is potentially misleading since the D/H ratio of water on Earth probably did not remain unchanged for the past 4.5 Gyr. Moreover, more recent measurements of other comets by [11] show that their $^{16}O/^{18}O$ ratios are consistent with the terrestrial values, especially among long period comets. Therefore, if water had extraterrestrial origins, it is more likely that a mixing of comets, the solar nebula and meteorites formed the present seawater.

Being only weakly bound to the Sun's gravity, comets in the Oort comet cloud are easily perturbed. Inner Oort cloud comets have aphelion Q < 20 000 AU, while outer Oort cloud comets have Q > 20 000 AU. At large distances from the Sun, the dominant perturbers of comet orbits are Giant Molecular Clouds (GMCs), passing nearby stars and the Galactic tides (see review by [12]). They can produce temporary increases the flux of comets entering the region of the planets – termed "comet showers" [13, 14].

The formation of a comet cloud depends on several factors.[1] The Sun's Oort cloud probably formed from planetesimels scattered by the giant planets [15-17]. However, we cannot generally apply the results of comet cloud formation simulations based on the Solar System's Oort cloud, given that the properties of giant planets are known to vary dramatically among nearby Sun-like stars [18]. One of the most important factors in determining the properties of giant planets is the metallicity of the birth cloud. More metal-rich systems are more likely to harbor giant planets [19]; reviewed by [20]. [21] successfully reproduced this relationship qualitatively with a model based on the core accretion theory of giant planet formation. Once we can successfully simulate the formation of giant planets in any planetary system, then it should be possible to determine the properties of the resulting comet cloud from simulations (a task beyond the scope of the present study).

The initial metallicity of a planetary system should also determine the availability of solid bodies that can be scattered into its Oort cloud [2]. Taken together, the observed correlation between metallicity and giant planet formation and the expected correlation between metallicity and planetesimal formation implies that there should be a strong correlation between metallicity and the properties of a comet cloud (e.g., its initial population). Determination of their precise relationship will require extensive simulations.

However, what can be said already, beginning with fundamental considerations of gravitational dynamics, is that at the earliest times, the orbits of cloud comets are influenced by the gravitational tidal field from its birth cloud/cluster and close encounters from nearby stars in the cluster [22]. Most stars are born in clusters, but most clusters suffer "infant mortality" [23], dissolving within a few million years. A cluster that survives infancy continues to dissolve as its stars are removed by tidal forces from encounters with GMCs [24], Galactic tides [25] and passages through spiral arms [26]. The cluster dissolution timescale should vary with location in a spiral galaxy; for example, a short dissolution timescale has been measured for the inner region of M51 [27], due to the high density of GMCs there [28]. In fact, the problem of the dynamical evolution of a star cluster is similar to that of an Oort cloud; both consist of many "test particles" influenced by the tidal field of a large mass distribution.

---

[1]Throughout this paper we refer to Oort clouds around other stars as comet clouds and to the comet cloud around the Sun as the Oort cloud.

The perturbations that a comet cloud experiences while it is within its birth cluster environment should have a significant effect on its dynamical evolution. Simulations show that the structure of the Sun's inner Oort cloud must have been sensitive to the initial density of its birth cluster [22, 29, 30]. A little explored topic is the effect of a binary companion on a comet cloud, whether it is a temporary one in a cluster environment [31] or a permanent one.

The effects of Galactic tides on the Sun's Oort cloud comets have been studied by many authors (e.g., [32-37] and others). Galactic tides are usually separated into radial, transverse and orthogonal (to the disk plane) components. At the Sun's distance from the Galactic center the orthogonal tide has been estimated to be about ten times more effective than the radial tide in perturbing Oort cloud comets into the inner Solar System [12, 34]. The orthogonal tide is the dominant perturber today and also over the long term [38]. [39] present observational evidence for the influence of the orthogonal tide from the observed distribution of the orbital elements of new comets (here, "new comets" refers to comets considered to be visiting the planetary region for the first time).

But while the orthogonal tide is predominant for the Sun's comets, as we will show, planar tides are not negligible. When the comet cloud is placed closer to the Galactic center these alone are sufficient to produce dramatic effects on their orbits.

In summary, the properties of a comet cloud depend on Galactic-scale factors in the following important ways. First, the radial disk metallicity gradient influences the initial population of comets in a comet cloud. Recent determinations of the metallicity gradient for the gas phase (or "zero age" objects) of the disk have converged on a value near $-0.07$ dex kpc$^{-1}$ [40-42], with a time derivative of 0.005 to 0.010 dex kpc$^{-1}$ Gyr$^{-1}$ [43]. Second, the disk surface densities of stars and GMCs increase steeply towards the Galactic center, as do the Galactic tides (see Section 3.1 below). As a result of these trends, comet clouds around stars closer to the Galactic center than the Sun will experience greater perturbations. Third, the frequency of spiral arm passages will vary depending on the distance from the corotation circle, where the spiral arm and stellar angular frequencies match.

## 1.2 SIMULATING COMET CLOUDS

To date, research on the formation and dynamical evolution of comet clouds has focused on the Solar System within its local Galactic environment. The orbits are followed numerically using the gravitational potential produced by a solar mass star imbedded in the local Galactic potential (i.e., which results from the Sun's location about 8 kpc from the Galactic center). The only exceptions have been studies of the impulsive effects of encounters between the Solar System and GMCs and nearby stars (neither of which is presently occurring) and also the changing orthogonal tide due to the Sun's vertical motion relative to the disk plane (e.g., [14]).

Studies of comet cloud dynamics have typically employed approximate analytical calculations or statistical analyses through Monte Carlo simulations. These approaches are understandable given the long orbital periods of the comets in relation to the orbital periods of the giant planets and the vast number of comets ($\sim 10^{12}$). Exact integrations of the full orbits of a large number of comets under the influence of one or more of the important perturbers are computationally expensive. Nevertheless, the approximate methods are useful for describing the dynamics of comets in certain restricted problems.

A completely self-consistent treatment of the Solar System's Oort cloud dynamics would have to begin with the four giant planets as they form in the protoplanetary disk and follow the planetesimals as the planets scatter some of them into the Oort cloud to begin their lives as comets (e.g., [44]). This initial distribution of the Oort cloud comets evolves over time as it experiences perturbations by passing stars and GMCs and the Galactic tides, sometimes injecting them back into the planetary region. The total population of Oort cloud comets declines over time and their distribution within the volume of space around the Sun will differ from the initial one [37]. Therefore, the distribution of comets in the Sun's Oort cloud inferred from observations of long period comets is just a snapshot in time of a continuously evolving system.

Clearly, the dynamics of the Oort cloud comets are complex and can't be exhaustively described in one study. In this paper we consider a restricted problem as a first step in removing the Oort cloud from its local solar and Galactic context and placing it in a more general setting. In particular, we will explore how the dynamics of a comet cloud depend on the planar Galactic tides[2] and also on stellar mass. Interactions with the planets, GMCs and nearby stars are not included in this first study, nor are the effects of different planetary architectures on the properties of a comet cloud. How comet clouds differ among the known extrasolar planets is an interesting question, but it is not one we will address in the present study. Exact integrations of a few dozen comet orbits should suffice to demonstrate, nevertheless, some qualitative aspects of their dynamics.

---

[2]Note, in the present study we are restricting our simulations to the Milky Way Galaxy's disk plane. Thus, the orthogonal disk tide is not relevant to our case, and the combined action of the radial and transverse tides in the disk plane might be better termed the "planar tides".

The paper is organized as follows. In Section 2.1, we present the analytic Galactic mass distributions and corresponding potentials that we use to integrate the orbits of the stars and their associated comets. In Sections 2.2 and 2.3 we describe the algorithm and its implementation for simulating comet orbits at different locations in the Galaxy and around stars of different masses. In Section 3 we present the results of our simulations. Finally, in Section 4, we suggest future lines of research, which should result in more realistic comet simulations.

## 2. Comet orbits with planar Galactic perturbations

The first step in simulating the dynamics of a comet cloud is to construct a model of the distribution of matter in the Milky Way Galaxy. This permits the calculation of the Galactic gravitational potential. A realistic potential would require accounting for the many irregularities and asymmetries in the mass distribution (e.g., the rotating triaxial bulge and the spiral arm pattern), leading to significant complications in the calculations. For this reason, in the present study we adopt a simple azimuthally symmetric Galactic model. A better determination of the matter distribution in the Galaxy and a more realistic model of its gravitational potential would be one way of improving on our simulations. However, as we are going to show, the precise determination of the Galactic force field is not necessary for the preliminary results of the present paper.

2.1 THE GALACTIC MASS DISTRIBUTION AND POTENTIAL

We built a simple but representative Galactic mass distribution with the following components.
1. We represent the giant black hole at the center of the Galaxy as a point with mass of $3.7 \times 10^6 M_\odot$ [45]. Its potential is given by $\Phi_{BH}$. Its contribution is negligible for calculating the comet orbits in the outer Galactic regions, but we include it nevertheless to show its effects on the radial tide in the central regions.

2. We treat the bulge and disk components together and employ the [46] potential form recommended by [47]. In our calculations, we adopted a four component Miyamoto-Nagai potential for the bulge and disk:

$$\Phi_{BD}(X,Y,Z) = -\sum_{n=1}^{4} \frac{GM_n}{\sqrt{X^2 + Y^2 + \left[a_n + \sqrt{b_n^2 + Z^2}\right]^2}}. \qquad (1)$$

We list the values of the $a_n$, $b_n$ and $M_n$ constants in Table 1. This formulation does not permit us to examine separately the potentials of the bulge and thin disk, but this disadvantage is outweighed by the greater efficiency it permits in the numerical computations compared to other formulations.

Originally [48], we had parameterized the mass distribution of the bulge as a Plummer sphere:

$$\rho_{BG}(r) = \left(\frac{3M_{BG}}{4\pi r_c^3}\right)\left(1 + \frac{r^2}{r_c^2}\right)^{-\frac{5}{2}},$$

with a corresponding potential:

$$\Phi_{BG}(r) = -\frac{GM_{BG}}{\sqrt{r^2 + r_c^2}},$$

with the following parameter values: $r_c = 420$ pc and $M_{BG} = 1.6 \times 10^{10}$ M$_\odot$ [47].

In addition, we had modeled the mass distribution of the thin disk with a two dimensional Freeman model [49], having an exponential surface mass density distribution:

$$\Sigma(r) = \Sigma_0 e^{-r/r_d},$$

where $\Sigma_0$ is the central disk surface mass density and $r_d$ is the disk's scale length. After a comparative analysis of different models (e.g., [50, 51]), we had adopted the following values for the constants: $\Sigma_0 = 492$ M$_\odot$ pc$^{-2}$, $r_d = 3.5$ kpc. These values are in agreement with the surface density at the solar distance of [52]. The corresponding potential for this mass distribution is:

$$\Phi_d(r) = -\pi G \Sigma_0 r \left[I_0(r/2r_d)K_1(r/2r_d) - I_1(r/2r_d)K_0(r/2r_d)\right],$$

where $I_n$ and $K_n$ are modified Bessel functions. We found this parameterization of the disk potential to be slow to integrate (and would be impractical in 3D), so we replaced them with the Miyamoto-Nagai parameterization in Eq. (1). We did, however, adjust the constants in Eq. (1) to match the potentials of the Plummer bulge and exponential disk models.

3. One representation of the matter distribution of the dark matter (DM) halo is a spherical pseudo-isotherm (PSISO):

$$\rho_{PSISO}(r) = \frac{\rho_0}{1+\left(\frac{r}{r_H}\right)^2}, \tag{2}$$

with $\rho_0$ the central density, $r$ is the Galactocentric distance and $r_H$ the scale length of the dark halo (following cosmological arguments made by [53], we chose $\rho_0 = 0.05$ M$_\odot$ pc$^{-3}$, and in order to recover a flat rotation curve, $r_H = 4.6$ kpc ). Eq. (2) represents the distribution of DM used in the old Schmidt's model [50]. We will consider more up to date models for the halo below. The PSISO distribution has the following potential (see [54]):

$$\Phi_{PSISO}(r) = -4\pi G \rho_0 r_H^2 \left[ 1 - \frac{r_H}{r}\arctan\left(\frac{r}{r_H}\right) - \frac{1}{2}\log\left(\frac{1+\left(\frac{r}{r_H}\right)^2}{1+\left(\frac{R_{vir}}{r_H}\right)^2}\right) \right] \tag{3}$$

where $R_{\text{vir}}$ is the extension of the virialized dark matter halo (this value however is not essential for our considerations because, when calculating the component of the force field associated with $\Phi_{PSISO}(r)$, $F_{PSISO} = -\nabla\Phi_{PSISO}$, $R_{\text{vir}}$ falls out in the derivative owing to Newton's first theorem).

Of these three components of the Galactic mass distribution, the DM halo is the most uncertain. In order to test the sensitivity of comet orbits to uncertainties in the mass distribution of the DM halo, we also consider two alternative formulations of the halo. One mass density distribution is the Navarro-Frenk-White (NFW) model [55]:

$$\rho_{NFW}(r) = \frac{\rho_0}{\frac{r}{r_H}\left(1+\frac{r}{r_H}\right)^2},$$

Its potential is given by

$$\Phi_{NFW}(r) = -4\pi G \rho_0 r_H^2 \left[ \frac{r_H}{r}\log\left(1+\frac{r}{r_H}\right) - \frac{1}{1+\frac{R_{vir}}{r_H}} \right].$$

The second is a modified pseudo-isotherm (MPSISO; [56]):

$$\rho_{MPSISO}(r) = \frac{\rho_0}{\left(1+\left(\frac{r}{r_H}\right)^2\right)^{3/2}}.$$

Its potential is given by

$$\Phi_{MPSISO}(r) = -4\pi G \rho_0 r_H^2 \left[ \frac{r_H}{r} \operatorname{arcsinh}\left(\frac{r}{r_H}\right) - \frac{1}{\sqrt{1+\left(\frac{R_{vir}}{r_H}\right)^2}} \right].$$

The two profiles are characterized by two different inner slopes with two extreme values. The profile of the NFW model has a slope value equal to –1, while the MPISO model has an inner slope equal 0. At this time it is unknown which of these is more realistic, even if there are some lines of evidence that suggest the value –0.4 [57, 58].

In order to choose suitable values for the parameters that appear within the two DM halo model alternatives to PSISO, we will follow the nice models proposed by [59]. In particular their favored model is A1 (with no exchange of angular momentum; see Table 1 of their paper). In a ΛCDM cosmology, with $H_0 = 70$ km s$^{-1}$ Mpc$^{-1}$ and $\Omega_0 = 0.3$ (due to the dark+ baryonic contribution) the virial mass of the DM halo is given by:

$$M_{vir} = \frac{4\pi}{3} \rho_{crit} \Omega_0 \delta_{th} R_{vir}^3,$$

where $\rho_{crit}$ is the critical density of universe and $\delta_{th}$ is the overdensity of a collapsed object in the "top-hat" collapse model ($\delta_{th} \approx 340$ for our cosmological model). Following [59], we chose a virial mass for the halo of $M_{vir} = 10^{12}$ M$_\odot$ and a concentration $C = [R_{vir} / r_H] = 17$ in the case of the NFW model. In this case, the scale density (at redshift z = 0; here and following we will neglect the correction due to the collapse redshift, $z_{col}$, at which the collected mass, $M_{vir}$, is collapsed in the clustering scenario, according to [55]) and the scale radius become, respectively: $\rho_0 = 0.012$ M$_\odot$ pc$^{-3}$ and $r_H = 15$ kpc. In the case of MPSISO the same mass is assumed with the values of $\rho_0 = 0.014$ M$_\odot$ pc$^{-3}$ (at z = 0) and $r_H = 13$ kpc, which correspond to $C = 20$.

The PSISO model we use is absolutely denser than both of these and has a scale radius smaller than the previous ones. It is in good agreement with the Schmidt model. We used it but explored also the other two alternatives. Its absolute slope value external to the solar circle might be a bit too low, but this region of the Galaxy is not relevant to our study (according to Newton's First Theorem). The PSISO model has the largest enclosed mass for values of $r$ beyond the solar circle. The MPSISO model yields a circular velocity of 196 km s$^{-1}$ at the solar circle, while the PSISO and NFW models yield velocities near 220 km s$^{-1}$ there.

Before describing our algorithm, it is helpful to visualize how the matter distributions and force fields vary with $r$. We begin by plotting in Fig. 1 the enclosed mass as a function of $r$ for the bulge, disk and DM halo (PSISO) Galactic components as well as their sum. Fig. 2 shows the circular velocity in the disk as a function of *r*. The velocity remains constant at large values of *r* in order to match the observed rotation curve.

In Fig. 3 we show how the Galactic gravitational force varies with *r*. The force field due to the bulge dominates only within 4 kpc of the Galactic center, whereas in the solar neighborhood the disk and halo are stronger than the bulge.

Of greater interest to us is the variation in radial tidal force, $F_{tidal} = \nabla F(r) = -\nabla^2 \Phi(r)$, which we show in Fig. 4. As expected, the radial tidal force increases towards the Galactic center. In our model, it reaches a maximum value at 500 pc from the center and then becomes negative (compressive) inside 300 pc. At *r* = 2 kpc the radial tidal force is about 23 times greater than in the solar neighborhood.

The fact that radial tides can also be negative (i.e., compressive instead of disruptive) is a natural aspect of gravity itself, which, however, is frequently overlooked. [60] first studied it in the context of disk shocking of star clusters. Later, [61] described compressive tides experienced by a disk galaxy falling into the core of a galaxy cluster. For a more extensive and analytic description applied to Galactic mass distributions see also [62].

We compare the radial and transverse tides in Fig. 5. Notice that they have very different functional dependencies on *r*, especially close to the Galactic center. The radial tide can be positive or negative, depending on the value of *r*, but the transverse tide is always negative. For values of *r* greater than about 3 kpc, the absolute values of the radial and transverse tides are comparable.

We also explored the variation of the tidal force with *r* using our other two formulations of the dark matter halo (NFW and MPSISO). Over the range of values of *r* we explore below (2 to 8 kpc), the NFW model gives stronger tides than the other two models. The MPSISO model gives slightly stronger tides than the PSISO model, but only for values of *r* less than about 5 kpc.

2.2 DESCRIPTION OF THE ALGORITHM

If we go from spherical to Cartesian galactocentric coordinates, through the substitution $r = \sqrt{X^2 + Y^2 + Z^2}$, then the Galactic gravitational potential at the position of a comet in Galactocentric coordinates $(X, Y, Z)$ can be represented as:

$$\Phi_{Gal}(X,Y,Z) = \Phi_{BH}(X,Y,Z) + \Phi_{BD}(X,Y,Z) + \Phi_D(X,Y,Z). \qquad (4)$$

Adding the gravitational potential of the star, $\Phi_*$, with Galactocentric position $(X_*, Y_*, Z_*)$ produced at $(X, Y, Z)$, gives:

$$\Phi_*(X - X_*, Y - Y_*, Z - Z_*) = -\frac{GM_*}{\sqrt{(X - X_*)^2 + (Y - Y_*)^2 + (Z - Z_*)^2}}. \quad (5)$$

Then the total potential, $\Phi_{Tot}$, becomes:

$$\Phi_{Tot}(X, Y, Z, X_*, Y_*, Z_*) = \Phi_{Gal}(X, Y, Z) + \Phi_*(X - X_*, Y - Y_*, Z - Z_*). \quad (6)$$

In the present study we are restricting our analysis to the Galactic planar tides, which act on a comet nucleus in the Galactic disk plane. In this first set of simulations we also neglect the small orthogonal excursions of the Sun (< 100 pc; [14]) and reduce the problem to a 2D one. By applying the usual equations of motion, $F_c = -\nabla \Phi_{Tot}$, to the comet we obtain the differential systems:

$$\begin{aligned}
\ddot{X}(t) &= -\frac{\partial \Phi_{Tot}}{\partial X}[X(t), Y(t), X_*(t), Y_*(t)]; \\
\ddot{Y}(t) &= -\frac{\partial \Phi_{Tot}}{\partial Y}[X(t), Y(t), X_*(t), Y_*(t)]; \\
X(0) &= R_g + R_c; \quad Y(0) = 0; \\
\dot{X}(0) &= 0; \quad \dot{Y}(0) = V_c,
\end{aligned} \quad (7)$$

where $(X(0), Y(0))$ and $V_c$ are, respectively, the initial position and speed of a comet, $R_c$ is its initial distance from the star, and $R_g$ is the initial distance of the star from the Galactic center.

The first step is to determine the star's orbit in time, $(X_*(t), Y_*(t))$. This is done in an analogous way through $F_* = -\nabla \Phi_{Gal}$, that is:

$$\begin{aligned}
\ddot{X}_*(t) &= -\frac{\partial \Phi_{Gal}}{\partial X}[X(t), Y(t)]; \\
\ddot{Y}_*(t) &= -\frac{\partial \Phi_{Gal}}{\partial Y}[X(t), Y(t)]; \\
X_*(0) &= R_g; \quad Y_*(0) = 0; \\
\dot{X}_*(0) &= 0; \quad \dot{Y}_*(0) = V_{t_*},
\end{aligned} \quad (8)$$

with $V_{t_*}$ the initial tangential velocity of the star (for our Sun $V_{t_*} \sim 220$ km s$^{-1}$ and $R_g \sim 8$ kpc). After numerical integration we obtain the trajectory of the comets in Galactocentric coordinates, which can be transformed into star-centric coordinates:

$$x(t) = X(t) - X_*(t); \quad y(t) = Y(t) - Y_*(t). \quad (9)$$

## 2.3 DESCRIPTION OF THE NUMERICAL METHOD AND 2D SIMULATIONS

The primary goal of the present study is to test whether the effects of Galactic planar tides alone, which are relatively unimportant at the Sun's location, become sufficiently strong at smaller Galactocentric distances to inject nearly parabolic comets towards the inner planetary region. Another goal is to study how comet orbits depend on stellar mass. We study these problems using numerical integrations of highly elliptic orbits of individual comets around stars moving on circular Galactic orbits but at different distances from the Galactic center (2, 4, 6 and 8 kpc).[3] Our second set of simulations treats a range of stellar masses (0.2, 0.8 and 2 $M_\odot$) at 8 kpc.

Comet orbits can become highly eccentric due to close encounters with giant planets, and traditional integration schemes can fail in reproducing efficiently these encounter phases. Galactic tides can perturb comets into an almost free-fall path (i.e., highly eccentric orbits for which perihelion passage can be challenging to integrate). Several efficient numerical integration schemes for these kinds of situations have been described in the literature (e.g., [63-65]). We will adopt one of them in our future simulations, especially when we include perturbations by the giant planets. In the present study, however, we still don't need these integration methods because there are no close encounters with planets to consider. We are only interested in the effects of the planar tides. These also can produce extremely eccentric encounters with the star, having perihelia of only a few dozens of AU, or less, and which are difficult to integrate. But this occurs in a region where the potential of the star is much greater compared to the Galactic field. Therefore, this segment of a comet's orbit near perihelion can be determined analytically with Kepler's laws, neglecting tidal or any other perturbations. The remaining orbital path, where the numeric error induced by high eccentricities is less and can be smoothed out with adaptive time steps methods, can be obtained, for instance, via a fourth-order Runge-Kutta integrator, as we have done. The overall integration procedure is as follows.

First, we integrated the orbit of the star in the Galactic potential. Next, we stored the star's orbit data and interpolated it with a Hermite polynomial for later use in the integration of the comet orbits.[4] This procedure saved time by requiring us to calculate a star's orbit only once for all the comets around a given star.

---

[3]We note that in the true Galactic potential elliptic orbits do not exist; also, in the present case we neglect the stochastic interactions with other stars and interstellar clouds. It would be more precise to regard planar stellar orbits as rosetta-like orbits.

We replaced the Runge-Kutta integration with a simple Keplerian analytic integration when the stellar force on a comet was 1000 times greater than the Galactic force. Under these conditions the Galactic force can be neglected. This occurred, for example, when a comet was within 80 AU of a one solar mass star 8 kpc from the Galactic center. This procedure saved a great deal of computation time, as the time step would have to be very small with the Runge-Kutta method to properly integrate the highly eccentric orbit of a comet very near the star.

Adjustment of the time steps is an important timesaving procedure, given the large variation in the speed of a comet with a highly eccentric orbit. We implemented adaptive time steps in the following way. From numerical experiments, we determined that the curvature of the stellar orbit and the relative velocity range between a comet and the star (determined from the comet's eccentricity) have the most direct influence on the change in the total energy. This tidal energy change is made up of a real energy change of the system due to its movement inside the Galactic mass distribution, plus numerical error growing in time in absolute value. Both are unknown quantities; however, minimizing the total energy change automatically leads also to numerical error minimization. The code uses these two parameters to set the time step size for a given tolerance in the total energy change of the orbit. Finally, from additional experiments, we established the maximum permitted total energy change, below which the orbits are indistinguishable for the duration of the integration. Once this limiting energy change tolerance was set for one comet, all the comets in the simulation were integrated with the same settings. The typical time step size for a comet around a one solar mass star at 8 kpc from the Galactic center was a few thousand years (the full range was 3 to 40 000 years).

We set the error tolerance for the integration of the star's orbit independently; we set the step size for the star's orbit at 1000 years, but preliminary tests show that a step size of 10 000 years probably would have been adequate.

We simulated the orbits of 48 comets, each with the same initial perihelion, $q$, of 2000 AU, and with different aphelia and different Galactic longitudes of the semi-major axis (see Table 2 for details). These 48 cases sample a wide range of the Sun's Oort cloud comets, and they are considered to represent some of the most distant comets still bound to the gravitational field of the Sun and perturbed by external Galactic forces. Nevertheless, they represent a plausible set of parameter values sampling the Sun's present Oort cloud in the Galactic plane [66-68].

---

[4] Hermite interpolation is a method particularly suited for dynamical data sets such as ours since it allows us to consider given derivatives at data points (i.e. here the star's and comet's velocities), as well as the data points themselves, i.e. it exploits the full dynamical information available.

Whether this set of initial conditions also holds for comet clouds around other stars in our Galaxy, which experienced different physical conditions and almost certainly had different dynamic and evolutionary histories, remains an open question. As long as we do not execute more sophisticated and realistic simulation models (and the aim of this paper is just to encourage such lines of research!), we will, however, continue to use this sample because it can give us at least a qualitative idea about the sensitivity of outer cloud comet orbits with changes either in the distance from the Galactic center or in the star's mass.

We present some of the results of our numerical simulations in Figs. 6 to 9. We show the results for the following cases: $r$ = 8, 6, 4, 2 kpc and stellar mass = 1 $M_\odot$ (Figs. 6 to 8), $r$ = 8 kpc and stellar mass = 0.2 $M_\odot$ (Fig. 9). We list the stellar orbital parameters in Table 3. We integrated each comet orbit for 100 Myr. This time limit was set by the available computing resources at the time we conducted our simulations, but it is sufficient to show the existence of the comet injection phenomenon.

When interpreting these figures, it is important to remember that the (x, y) reference frame on which the comet orbits are plotted is rotating and revolving around the Galactic center (e.g., with a 224 Myr period in the case of our Sun at 8 kpc). An observer fixed to the (x, y) frame "sees" the Galactic center moving around him. As the (arbitrary) initial condition for each case, we set the x-axis to correspond to the star-Galactic center axis (Galactic longitude of 0°). But this alignment is true only at t = 0. For instance, consider the major axis of a comet orbit which lies on the x-axis (parallel to the radial Galactic coordinates), with its minor axis orthogonal to it, then, after a quarter of the Galactic orbital period the situation is inverted; it is the minor axis that points toward the Galactic center, while the major axis is parallel to Galactic longitude 0°, but is displaced from it. So, there is no orbital axis pointing towards any particular Galactic direction over the full timespan of the simulation.

Finally, we should note that the results of the simulations presented in Figs. 6 to 9 are based on the disk+bulge potential given by Eq. 1 and the PSISO halo potential given by Eq. 3. We repeated each simulation using the NFW and MPSISO halo potentials in place of the PSISO potential in order to test the sensitivity of the comet orbits to the precise form of the halo potential. We discuss the outcomes of these additional simulations below.

# 3. Discussion

## 3.1 RESULTS

Our 2D simulations confirm that the planar tides are more important for comet clouds located closer to the Galactic center. They also verify that the larger comet orbits are more strongly perturbed by the Galactic tides than the smaller ones. What's more, the perihelia of the perturbed comets are reduced to regions far inside their initial values. For the $r = 8$ kpc simulations, two comets among those with the largest aphelia had their perihelia reduced to within 100 AU of the star. For the $r = 6$ kpc simulations, the smallest perihelia are less than 2 AU, and they are less than a quarter of an AU for the $r = 2$ kpc simulations.

At $r = 2$ kpc the comet clouds are well inside the bulge. The planar tides are so strong there that the Galactic Roche limit is inside the comet cloud. Several comets with aphelia > 100 000 AU were lost from the comet cloud by the end of the simulation. In addition, comets with smaller aphelia had their perihelia significantly reduced.

To better understand the results of our simulations recall, as outlined at the end of subsection 3.1, that the figures are not shown in the Galactic radial/longitudinal reference frame, but in the stellar reference frame that moves around the Galactic center. First, consider those orbits that appear horizontal in Figs. 6-9. We set as an initial condition that their major axes are aligned exactly with the Galactic longitude 0° only at time t = 0, but one can consider this alignment approximately valid for several million of years. The effect acting on these orbits works in a way similar to the Moon-Earth tides: along the radial coordinate one has disruptive tides which produce the classical ocean bulges, while along the transverse and orthogonal coordinates the tides are compressive and produce a tiny polar flattening. There is a difference, however, since we are considering, not a solid object, but loosely orbiting particles. The semi-minor axes tend to be increased despite the presence of the compressive transverse tidal forces. This happens for the following reasons. Due to the positive (disruptive) radial Galactic tide the aphelia are increased, making comets even more loosely bound to the stellar gravitational field, and this consequently leads to the increase of their minor axes too since, because of their (nearly) unaltered (at this time) $Y^*$-component kinetic energy, they are allowed to make a greater excursion along the semi-minor axis. The transverse tides are negligible or less dominant (at this time) since the excursion along the semi-minor axis remains nevertheless small compared to that of the major axis. Therefore, one has a sort of "inflating" effect of the orbit. At perihelion the comet is attracted toward the Galactic center, so it increases its perihelion too. The net effect is a shift of the perihelia <u>outward</u> from the center of the stellar system.

Vice versa, for those comets that have at t = 0 their major axes aligned 90° in Galactic longitude. In this case, the radial tides are negligible or less dominant, but the transverse tides are at work, and these are always negative (compressive)! Here, the aphelia tend to be shortened, the orbits are "squeezed" and the net effect is to shift perihelia towards the center of the stellar system. The reduction of a comet's perihelion in this case is caused by the transverse tides (this however does not mean that there isn't any radial dependence, since the intensity of the transverse tides is radially dependent too).

More generally, it is the existence of compressive tides that is responsible for the cometary injection phenomenon. Note that, despite limiting ourselves to a 2D simulation, this explains also why the orthogonal tides perturb Oort cloud comets into the solar system, as some authors outline in the works mentioned above. They demonstrated (possibly without being aware of it) the above-mentioned disk shocking effect of [60] for a comet cloud instead of a star cluster.

For those orbits aligned to other longitudes, one has a superposition of the two effects. Moreover, since the comet cloud is moving around the Galactic center, the effects of the radial and transverse tides are therefore continuously exchanged every quarter of the star's Galactic orbit. The final result is a complex mixture of these non-linear effects modulated in time, which cannot be separated from each other. The shift of the comet perihelia from far (close) to close (far) to (from) the star occurs during only one quarter of the stellar Galactic orbit, and then the effect switches the next quarter of the orbit, and so on. The apparently different evolution of the orbits in the figures with major axes of different angles can be misleading. The first case reproduces "thick" orbits and the other "thin" ones for the reasons explained above, but this is so only because of the different initial orientation at t = 0 of the orbital axis, which determines if the perihelia are moving outward or inward first: in the first case the perihelia move outward but then return back near the 2000 AU initial value of $q$, and in the second case they move inward first below the 2000 AU value, but also tend to return.

In summary, the Galactic planar tides provoke an oscillatory inward and outward movement of the comet perihelia linked to the periodic Galactic orbit. For our model at 8 kpc this "oscillation band" is not broad enough to "touch" the planetary region, but at 6 kpc it is.

In order to have a more statistically rigorous description of the change in perihelion distances of the comets, we counted the number of perihelion *transits* into the region between 0 and 3000 AU (i.e., if a single comet passes n times it will be counted n times, etc). Figure 10 shows a cumulative count of the comet perihelia into this inner region for the four different simulations at 8, 6, 4 and 2 kpc from the galactic center. The peak in the distribution for 8 kpc occurs near 2000 AU, causing the cumulative distribution to be steepest at this point.

At first sight the difference between the 8 kpc and 6 kpc case is hardly discernible, but the innermost region of the 6 kpc cloud is filled up slightly more. The comet clouds for 4 and 2 kpc from the galactic center instead show how the galactic tides clearly induce a 'flattening' of the cumulative distributions: the perihelia have been scattered towards the inner as well as the outer zones. Injection as well as depletion forces are at work. The injection phenomenon is highlighted by the populating of the innermost regions for the clouds nearest to the galactic center. Depletion is also evident as a reduction in the total number perihelion transits inside 3000 AU (599, 591, 548, 408 transits for 8, 6, 4, 2, kpc respectively). Overall, the 6 kpc distance from the galactic center seems to represent a limit beyond which dramatic changes begin to take place in the comet cloud dynamic due to galactic planar tides only.

For the simulations with different values of the star mass, the results are similar to changing the Galactocentric radius. For instance, in Fig. 9 we show the simulations for $r = 8$ kpc and stellar mass = 0.2 $M_\odot$. They appear qualitatively similar to the simulations for $r = 4$ kpc and stellar mass = 1 $M_\odot$. The smallest perihelion for this case is within 0.004 AU of the star, which would result in the destruction of the comet! As expected, the simulations for $r = 8$ kpc and stellar mass = 2 $M_\odot$ showed less perturbed orbits.

Simulations employing either the NFW or MPSISO alternative descriptions of the halo potential yielded qualitatively similar results to those shown in Figs. 6 to 9; viewed on the largest scales, the differences in the orbits are barely perceptible. In particular, our findings that the Galactic tides move the comet perihelia to the planetary region for $r \leq 6$ kpc and that the comet orbits for stellar mass = 0.2 $M_\odot$ at 8 kpc are similar to those for stellar mass = 1 $M_\odot$ at 4 kpc are unchanged. In other words, the phenomenon of comet injection into the planetary region is largely independent of the particular halo model adopted (so long as the halo model agrees with the observed Galactic rotation curve).

That said, we did find some differences using the alternative halo potentials. The comet orbits differ most near their perihelia and also for the small $r$ and small stellar mass cases, where the orbits are more chaotic. Of the three halo potentials, NFW produces the strongest tides. For the $r = 8$ kpc, stellar mass = 1 $M_\odot$ case, the NFW potential yielded a minimum perihelion value of 30 AU, compared to 75 AU for the PSISO case. While the NFW potential yielded the smallest perihelia in the linear, non-chaotic regimes, the outcomes for individual comets were not so predictable in the chaotic regimes.

We also ran some simulations with the halo potential "switched off." First, we show the effect on the Sun's orbit in Fig. 11. There are two cases: in one, the Sun continues orbiting with the current speed which develops into an elongated rosetta-like path towards the outer regions of the Galaxy. In the second, the Sun's velocity has been rescaled (161 km s$^{-1}$) in order to maintain a circular orbit at 8 kpc from the galactic center. We show the resulting comet orbits for the two cases in Fig. 12 . Fig.12a is obtained due to a change of orbit to the external regions of the halo-less Galaxy where, as expected (galactic tides less effective) the cometary injection is weakened, while Fig.12b shows the orbits one gets maintaining the constant distance of 8 kpc from the Galaxy center. Compare these with Fig. 7a. In particular Fig. 12b shows that, since there have been no major dynamic changes at these galactic distances, the tidal effects of the dark halo are less important than those of the bulge and disk, though not negligible (confirming also the tidal force functions shown in Fig.4).

In summary, while the overall results of our simulations do not depend on the precise form of the dark matter halo, nevertheless, the details are sensitive to it. This is a remarkable link between the structure of the Milky Way at the largest scales and the orbits of comets visiting the planetary region.

## 3.2 SUGGESTIONS FOR FUTURE RESEARCH

As we noted above, ours is a preliminary study focusing on two specific questions: how do the effects of the Galactic planar tides on a comet cloud change when the location and mass of a star are varied from the solar circumstances? We began to answer these questions in the present work with exact numerical simulations restricted to 2D, a small number of comets and a short time span. It is straightforward to remove any or all these restrictions, but the expanded simulations will require additional computational resources. Computations of comet cloud comet orbits are especially amenable to parallelization. Since each comet orbit is independent of the others, their orbits can be integrated on parallel CPUs. The only requirement is that the star's orbit in the Galaxy be integrated first.

The most important consequence of adding a third dimension to the simulations is to introduce the motion of a star and its comets perpendicular to the Galactic disk. Proper integration of their orbits will require smaller time steps, since the period for the vertical oscillation of the Sun relative to the disk is shorter than its period of revolution around the Galactic center. This will result in an orthogonal tide. The effect of the orthogonal tide is to increase the eccentricities of comets with significant inclinations relative to the Galactic plane [69]; this is the "squeezing" effect we mentioned in subsection 3.1. Over time, this tide will deplete the comet cloud of those comets with high inclinations.

Comet orbits can be "reshuffled" to repopulate the volume of a comet cloud following close encounters with stars and GMCs. They can be included in a simulation as short-lived stochastic encounters. The frequency of such encounters will be greater at smaller Galactocentric radii.

If the effects of passages through spiral arms are to be included in the simulations, then it will be necessary to build a more realistic Galactic potential than the one we employed. A nuclear bar and spiral arms must be added. One possible representation of the spiral arm potential is the one employed by [70] to simulate the orbits of stars in the Galactic disk.

Inclusion of perturbations from the four giant planets in the Solar System will result in accelerated loss of comets from the Oort cloud as well as reduction in the aphelia of those comets remaining bound to the Sun. Analysis of the "half-life" of a comet cloud under the influence of the planets and the other perturbers will require populating the initial comet cloud with a large number of comets distributed in a way that is consistent with its formation. At any time, the dynamics and structure of the comet cloud will depend on the details of its formation and on its prior dynamical history.

Generalizing the Sun's Oort cloud to other planetary systems will require modeling how the formation of giant planets varies with time, Galactic location and the mass of the star. Some recent progress has been made in understanding how the formation of giant planets depends on initial metallicity and stellar mass (e.g., [21]). It is already clear that relating the properties of a comet cloud to the properties of its host star and any accompanying giant planets will require Monte Carlo simulations of many cases.

The ultimate goal of this project is to simulate the formation and subsequent evolution (for at least a few Gyr) of a comet cloud anywhere in the Galaxy formed at anytime during its virialized phase. This will permit us to study the variation of the comet flux into the planetary region of a star statistically and thus further refine the boundaries of the GHZ.

## 4. Conclusions

We have simulated the orbits of comet clouds in 2D around stars having different Galactic locations and masses. We confirm that the Galactic planar tides are much more effective at perturbing comets into the planetary region for stars closer to the Galactic center than they are at the Sun's location. The Galactic tides alone are unable to send the Oort cloud comets into the Sun's planetary region in 100 Myrs, but they could if it were only about 2 kpc closer to the Galactic center. In addition, we find that comets around stars of smaller mass are more easily perturbed compared to the Sun's comets. The tidal perturbations of comets around a star of one solar mass at a Galactocentric radius of 4 kpc are comparable to those of comets around a 0.2 solar mass star at 8 kpc. The overall conclusions of our study are not changed if the dark matter halo potential is altered slightly, but it is important to calibrate it accurately with observations.

These preliminary results indicate that low mass stars that have planetary architectures similar to ours and are closer to the Galactic center than we are will have their outermost comets perturbed into the planetary region on a timescale of a few hundred Myr or less. This produces higher fluxes of comets in the circumstellar habitable zones of the stars, potentially leading to a higher frequency of impact-induced extinction events. Combined with the increased threat from supernovae and other hazards in the inner Galaxy [2], the results from the present study help to further refine the inner boundary of the GHZ.

## Acknowledgements

We wish to thank Prof. Vittorio Vanzani for his professional contribution to our discussions and Dr. Daniele Bindoni, who offered us his expertise on dark matter haloes.

## References

[1] Lineweaver CH, Fenner Y, Gibson BK. The Galactic habitable zone and the age distribution of complex life in the milky way. Science 2004; 303: 59-62.

[2] Gonzalez G. Habitable zones in the universe. Origins Life Evol Bios 2005; 35: 555-606.

[3] Sepkoski JJ. Patterns of Phanerozoic extinction: A perspective from global databases. In: Walliser OH, Ed. Global events and event stratigraphy, Springer, Berlin 1995; pp. 35-51.

[4] Brownlee D. et al. Comet 81P/Wild 2 under a microscope. Science 2006; 314: 1711-16.

[5] Matrajt G, Brownlee D, Sadilek M, Kruse L. Survival of organic phases in porous IDPs

during atmospheric entry: A pulse-heating study. Meteoritics 2006; 41: 903-11.

[6]   Chyba CF. The cometary contribution to the oceans of primitive Earth. Nature 1987; 330: 632-5.

[7]   Eberhardt P, Reber M, Krankowsky D, Hodges RR. The D/H and $^{18}O/^{16}O$ ratios in water from comet P/Halley. Astron Astrophys 1995; 302: 301-16.

[8]   Bockelée-Morvan D, Gautier D, Lis DC, Young K, Keene J, Phillips T, Owen T, Crovisier J, Goldsmith PF, Bergin EA, Despois D, Wooten A. Deuterated water in comet C/1996 B2 (Hyakutake) and its implications for the origin of comets. Icarus 1998; 133: 147-62.

[9]   Meier R, Owen TC, Matthews HE, Jewitt DC, Bockelée-Morvan D, Biver N, Crovisier J, Gautier D. A determination of the HDO/H2O ratio in comet C/1995 01 (Hale-Bopp). Science 1998; 279: 842-4.

[10]  Genda H, Ikoma M. Origin of the ocean on the Earth: Early evolution of water D/H in a hydrogen-rich atmosphere. Icarus 2008; 194: 42-52.

[11]  Biver N, Bockelée-Morvan D, Crovisier J, Lecacheux A, Frisk U, Hjalmarson Å, Olberg M, Florén H-G, Sandquist A, Kwok S. Submillimeter observations of comets with Odin: 2001–2005. Plan and Space Science 2007; 55: 1058-68.

[12]  Morbidelli A. Origin and dynamical evolution of comets and their reservoirs. arXiv: 0512.0256. 9 December 2005: Available from http://arxiv.org/abs/astro-ph/0512256.

[13]  Heisler J, Tremaine S, Alcock C. The frequency and intensity of comet showers from the Oort cloud. Icarus 1987; 70: 269-88.

[14]  Matese JJ, Whitman PG, Innanen KA, Valtonen MJ. Periodic modulation of the Oort cloud comet flux by the adiabatically changing galactic tide. Icarus 1995; 116: 255-68.

[15]  Fernández JA. The formation of the Oort cloud and the primitive Galactic environment. Icarus 1997; 129: 106-19.

[16]  Higuchi A, Kokubo E, Mukai T. Scattering of planetesimals by a planet: Formation of comet cloud candidates. Astron J 2006; 131: 1119-29.

[17]  Levison HF, Duncan MJ, Dones L, Gladman BJ. The scattered disk as a source of Halley-type comets. Icarus 2006; 184: 619-33.

[18]  Butler RP, Wright JT, Marcy GW, Fischer DA, Vogt SS, Tinney CG, Jones HRA, Carter BD, Johnson JA, McCarthy C, Penny AJ. Catalog of nearby exoplanets. Astrophys J 2006; 646: 505-22.

[19]  Fischer DA, Valenti J. The planet-metallicity correlation. Astrophys J 2005; 622: 1102-

17.

[20]     Gonzalez G. The chemical compositions of stars with planets: A review. Pub Astron Soc Pac 2006; 118: 1494-1505.

[21]     Ida S, Lin DNC. Dependence of exoplanets on host stars' metallicity and mass. Prog Theor Phys Suppl 2005; No 158: 68-85.

[22]     Brasser R, Duncan MJ, Levison HF. Embedded star clusters and the formation of the Oort cloud. Icarus 2006; 184: 59-82.

[23]     Lada CJ, Lada EA. Embedded clusters in molecular clouds. Ann Rev Astron Astrophys 2003; 41: 57-115.

[24]     Gieles M, Portegies Zwart SF, Baumgardt H, Athanassoula E, Lamers HJGLM, Sipior M, Leenaarts J. Star cluster disruption by giant molecular clouds. Mon Not R Astron Soc 2006; 371: 793-804.

[25]     Lamers HJGLM, Gieles M, Bastian N, Baumgardt H, Kharchenko NV, Portegies Zwart S. An analytical description of the disruption of star clusters in tidal fields with an application to galactic open clusters. Astron Astrophys 2005; 441: 117-29.

[26]     Gieles M, Athanassoula E, Poregies Zwart SF. The effect of spiral arm passages on the evolution of stellar clusters. Mon Not R Astron Soc 2007; 376: 809-19.

[27]     Gieles M, Bastian N, Lamers HJGLM, Mout JN. The star cluster population of M51. III. Cluster disruption and formation history. Astron Astrophys 2005; 441: 949-60.

[28]     Lamers HJGLM, Gieles M. Star clusters in the solar neighborhood: A solution to Oort's problem. In: de Koter A, Smith L, Waters R, Eds. Mass loss from stars and the evolution of stellar clusters, Astron Soc Pac, San Francisco, 2008; p. 367-79.

[29]     Fernández JA, Brunini A. The buildup of a tightly bound comet cloud around an early sun immersed in a dense galactic environment: Numerical experiments. Icarus 2000; 145: 580-90.

[30]     Kaib NA, Quinn T.: 2008, The formation of the Oort cloud in open cluster environments. Icarus 2008; 197: 221-38.

[31]     Malmberg D, De Angeli F, Davies MB, Church RP, Mackey D, Wilkinson MI. Close encounters in young stellar clusters: Implications for planetary systems in the solar neighborhood. Mon Not R Astron Soc 2007; 378: 1207-16.

[32]     Byl J. Galactic perturbations of nearly-parabolic cometary orbits. Moon Planets 1983; 29: 121-37.

[33]     Byl J. The effect of the galaxy on cometary orbits. Earth Moon Plan 1986; 36: 263-73.

[34]    Heisler J, Tremaine S. The influence of the galactic tidal field on the Oort comet cloud. Icarus 1986; 65: 13-26.

[35]    Torbett MV. Dynamical Influence of Galactic tides and molecular clouds on the Oort cloud of comets. In: Smoluchowski R, Bahcall JN, Matthews MS, Eds. The galaxy and the solar system, Univ Arizona Press, Tucson 1986; pp. 147-172.

[36]    Matese JJ, Whitman PG. A model of the galactic tidal interaction with the Oort comet cloud. Celest Mech Dyn Astron 1992; 54: 13-35.

[37]    Fouchard M, Froeschlé C, Valsecchi G, Rickman H. Long-term effects of the galactic tide on cometary dynamics. Cel Mech Dyn Astron 2006; 95: 299-326.

[38]    Heisler J. Monte carlo simulations of the Oort comet cloud. Icarus 1990; 88: 104-21.

[39]    Matese JJ, Lissauer JJ. Perihelion evolution of observed new comets implies the dominance of the galactic tide in making Oort cloud comets discernable. Icarus 2004; 170: 508-13.

[40]    Kovtyukh VV, Wallerstein G, Andrievsky SM. Galactic cepheids. I. Elemental abundances and their implementation for stellar and galactic evolution. Pub Astron Soc Pac 2005; 117: 1173-81.

[41]    Luck RE, Kovtyukh VV, Andrievsky SM. The distribution of the elements in the galactic disk. Astron J 2006; 132: 902-18.

[42]    Lemasle B, Francois P, Bone G, Mottini M, Primas F, Romaniello M. Detailed chemical composition of galactic cepheids. A determination of the galactic abundance gradient in the 8-12 kpc region. Astron Astrophys 2007; 467: 283-94.

[43]    Maciel WJ, Lago LG, Costa RDD. An estimate of the time variation of the abundance gradient from planetary nebulae. III. O, S, Ar and Ne: A comparison of PN samples. Astron Astrophys 2006; 453: 587-93.

[44]    Brasser R, Duncan MJ. An analytical method to compute comet cloud formation efficiency and its application. Cel Mech Dyn Astron 2008; 100: 1-26.

[45]    Ghez AM, Salim S, Hornstein SD, Tanner A, Lu JR, Morris M, Becklin EE, Duchêne, G.: 2005, Stellar orbits around the galactic center black hole. Astrophys J 2005; 620: 744-57.

[46]    Miyamoto M, Nagai R. Three-dimensional models for the distribution of mass in galaxies. Pub Astron Soc Japan 1975; 27: 533-43.

[47]    Flynn C, Sommer-Larsen J, Christensen PR. Kinematics of the outer stellar halo. Mon Not R astron Soc 1996; 281: 1027-32.

[48] Masi M, Secco L, Vanzani V. Dynamical effects of the galaxy on the Oort's cloud. Mem SA It 2003; 74: 494-5.

[49] Freeman KC.: 1970, On the disks of spiral and S0 galaxies. Astrophys J 1970; 160: 811-30.

[50] Schmidt M. Models of the mass distribution of the galaxy. In: van Woerden H et al. Eds. The milky way galaxy, D Reidel Publishing Comp, Dordrecht 1985; pp. 81-84.

[51] Van der Kruit PC. Surface photometry of edge-on spiral galaxies. V. The distribution of luminosity in the disk of the galaxy derived from the Pioneer 10 background experiment. Astron Astrophys 1986; 157: 230-44.

[52] Kuijken K, Gilmore G. The galactic disk surface mass density and the galactic force $K(z)$ at $Z = 1.1$ kiloparsecs. Astrophys J 1991; 367: L9-L13.

[53] Bullock JS, Kolatt TS, Sigard Y, Sommerville RS, Kravtsov AV, Klypin AA, Primacke JR, Dekel A. Profiles of dark haloes: evolution, scatter and environment. Mon Not R Astron Soc 2001; 321: 559-75.

[54] Binney J, Tremaine S. Galactic dynamics. Princeton University Press: Princeton 1987, p. 36.

[55] Navarro JF, Frenk CS, White SDM. A universal density profile from hierarchical clustering. Astrophys J 1997; 490: 493-508.

[56] Spano M, Marcelin M, Amram P, Carignan C, Epinat B, Hernandez O. GHASP: An H$\alpha$ kinematic survey of spiral and irregular galaxies. V. Dark matter distribution in 36 nearby spiral galaxies. Mon Not R Astron Soc 2008; 383: 297-316.

[57] Binney JJ, Evans NW. Cuspy dark matter haloes and the Galaxy. Mon Not R Astron Soc 2001; 327: L27-L31.

[58] Merrifield MR. The galactic bar. In: Clemens D, Brainerd T, Shah R, Eds. Milky way surveys: The structure and evolution of our galaxy, ASP Conf. Series, San Francisco, 2004; pp. 289-303.

[59] Klypin A, Zhao H, Somerville RS. ΛCDM-based models for the milky way and M31. I. Dynamical models. Astrophys J 2002; 573: 597-613.

[60] Ostriker JP, Spitzer L, Chevalier RA. On the evolution of globular clusters. Astrophys J 1972; 176: L51-L56.

[61] Valluri M. Compressive tidal heating of a disk galaxy in a rich cluster. Astrophys J 1993; 408: 57-70.

[62] Masi M. On the compressive radial tidal forces. Am J Phys 2007; 75: 116-24.

[63]   Hut P, Makino J, McMillan S. Building a better leapfrog. Astrophys J 1995; 443: L93-L96.

[64]   Levison HF, Duncan MJ. Symplectically integrating close encounters with the sun. Astron J 2000; 120: 2117-23.

[65]   Breiter S, Fouchard M, Ratajczak R, Borczyk W. Two fast integrators for the galactic tide effects in the Oort cloud. Mon Not R Astron Soc 2007; 377: 1151-62.

[66]   Fernández JA. Evolution of comet orbits under the perturbing influence of the giant planets and nearby stars Icarus 1980; 42: 406-21.

[67]   Weissman PR. Dynamical Evolution of the Oort cloud. In: Carusi A Ed. Dynamics of Comets; Their Origin and Evolution, D Reidel Pub Co, Dordrecht 1985: pp. 87-96.

[68]   Remy F, Mignard F. Dynamical evolution of the Oort cloud. I. A monte carlo simulation. Icarus 1985; 63: 1-3-.

[69]   Wiegert P, Tremaine S. The evolution of long-period comets. Icarus 1999; 137: 84-121.

[70]   Lépine JRD, Acharova IA, Mishurov Yu N. Corotation, stellar wandering, and the fine structure of the Galactic abundance pattern. Astrophys J 2003; 589: 210-6.

TABLE I

Parameter values for Eq. 1

| Parameter | Value |
|---|---|
| $a_1$ | 80 |
| $a_2$ | 500 |
| $a_3$ | 4600 |
| $a_4$ | 9000 |
| $b_1$ | 95 |
| $b_2$ | 200 |
| $b_3$ | 400 |
| $b_4$ | 200 |
| $M_1$ | $2.4 \times 10^9$ |
| $M_2$ | $18.7 \times 10^9$ |
| $M_3$ | $40.2 \times 10^9$ |
| $M_4$ | $3.4 \times 10^9$ |

TABLE II

Initial conditions of comet simulations

| Cases | Q (AU) | a (AU) | e | Period (Myr) |
|---|---|---|---|---|
| 1-8 | 40 000 | 21 000 | 0.9047 | 3.0 |
| 9-16 | 60 000 | 31 000 | 0.9355 | 5.5 |
| 17-24 | 80 000 | 41 000 | 0.9512 | 8.3 |
| 25-32 | 100 000 | 51 000 | 0.9608 | 11.5 |
| 33-40 | 120 000 | 61 000 | 0.9672 | 15.1 |
| 41-48 | 140 000 | 71 000 | 0.9718 | 19.0 |

*Note:* Within each of the six groups, the eight cases correspond to the following initial Galactic longitudes of the comet orbits: 0, 45, 90, 135, 180, 225, 270 and 315 degrees. The second column lists initial comet orbit aphelion values; all cases have an initial perihelion value of 2000 AU. The last column lists the comet's orbital period around a solar mass star in the absence of the Galactic potential. The initial conditions in this table apply to all four values of $r$ adopted in the simulations: 2, 4, 6 and 8 kpc.

TABLE III

Orbital parameters of the simulated stars

| r (kpc) | Stellar orbital period (Myr) |
|---|---|
| 2 | 57 |
| 4 | 125 |
| 6 | 173 |
| 8 | 224 |

*Note:* The first column lists the value of the star's Galactocentric distances. The second column gives its orbital period in the disk plane.

FIGURE CAPTIONS

*Figure 1*. Enclosed mass plotted against Galactocentric radius, *r*, for the bulge, disk and dark matter halo components. The components have the same representations in Figs. 2 to 4.

*Figure 2*. Circular velocity plotted against *r*.

*Figure 3*. Gravitational force plotted against *r*.

*Figure 4*. Radial tidal force plotted against *r*. These plots also include the contribution from the central black hole, indicated by long-dashed curves.

*Figure 5*. Radial and transverse tides plotted against *r*.

*Figure 6*. The full orbits of the 48 comets in our simulations with *r* = 8, 6, 4, 2 kpc in panels a, b, c, d, respectively, and star mass = 1 $M_\odot$. Each comet orbit is plotted in color depending on its size, going from light blue for the largest orbits to red for the smallest. The x and y axes are star-centric; see text for additional discussion.

*Figure 7*. Same as Fig. 6 but showing only the regions from -1500 to 1500 AU in x and y.

*Figure 8*. Same as Fig. 6 but showing only the innermost regions from -150 to 150 AU in x and y.

*Figure 9*. Simulation of comet orbits for *r* = 8 kpc and star mass = 0.2 $M_\odot$. All else as in Fig. 6a.

*Figure 10*. Cumulative count of the number of comet transits inside a radius of 3000 AU for *r* = 8, 6, 4, 2 kpc.

*Figure 11*. Sun's orbit if halo potential is suddenly "switched off" but keeps same orbital speed (blue). The 8 kpc circular orbit is shown in red with halo potential "on" (or halo off with orbital speed set to 161 km s$^{-1}$).

*Figure 12*. a) Comet orbits in region from -1500 to 1500 AUs in x and y for case with halo potential switched off with present orbit speed (blue orbit in Fig. 11). b) Comet orbits for red solar orbit in Fig. 11. Compare to Fig. 7a.

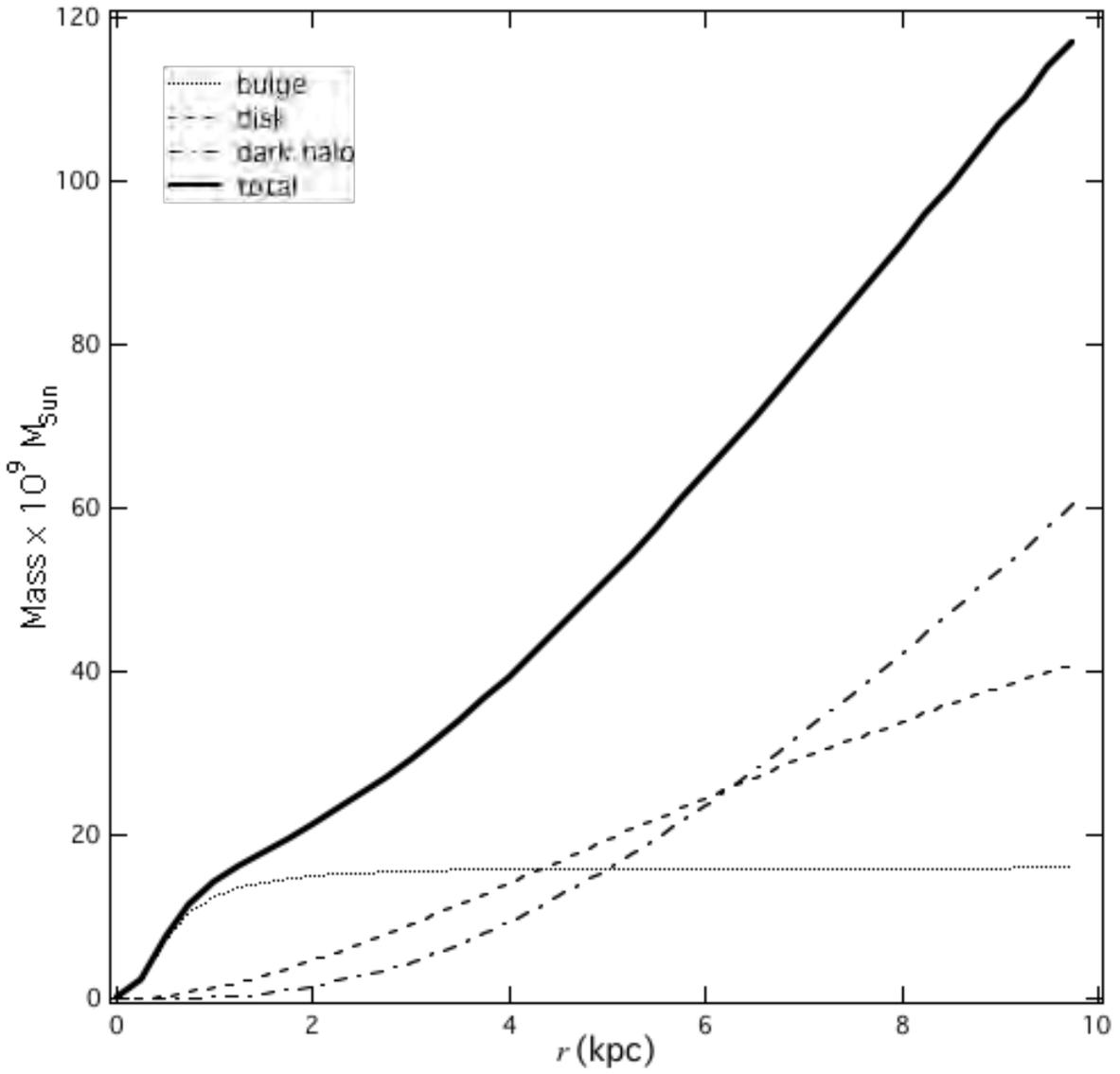

Figure 1: enclosed mass vs *r*

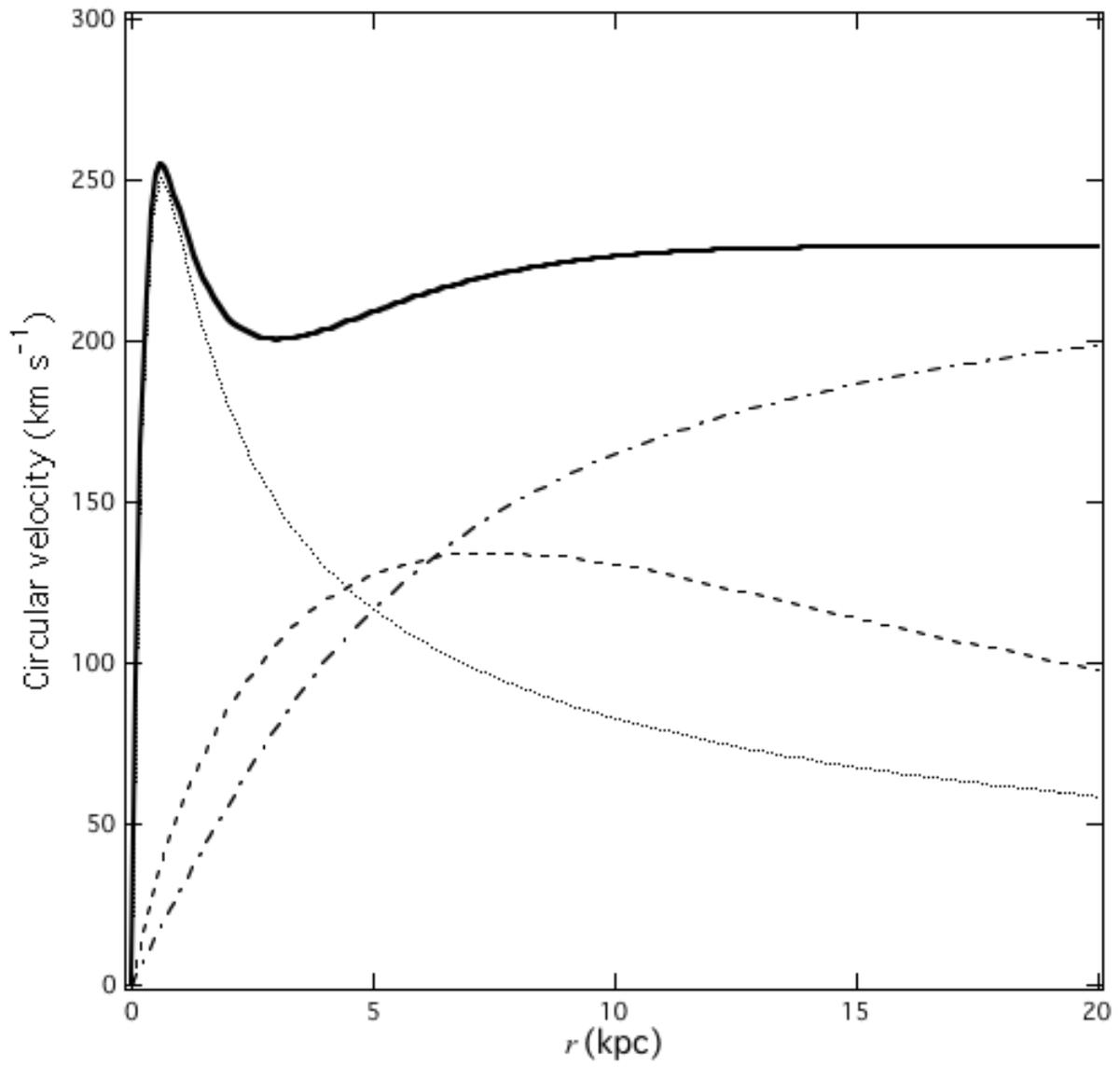

Figure 2: circular velocity vs *r*

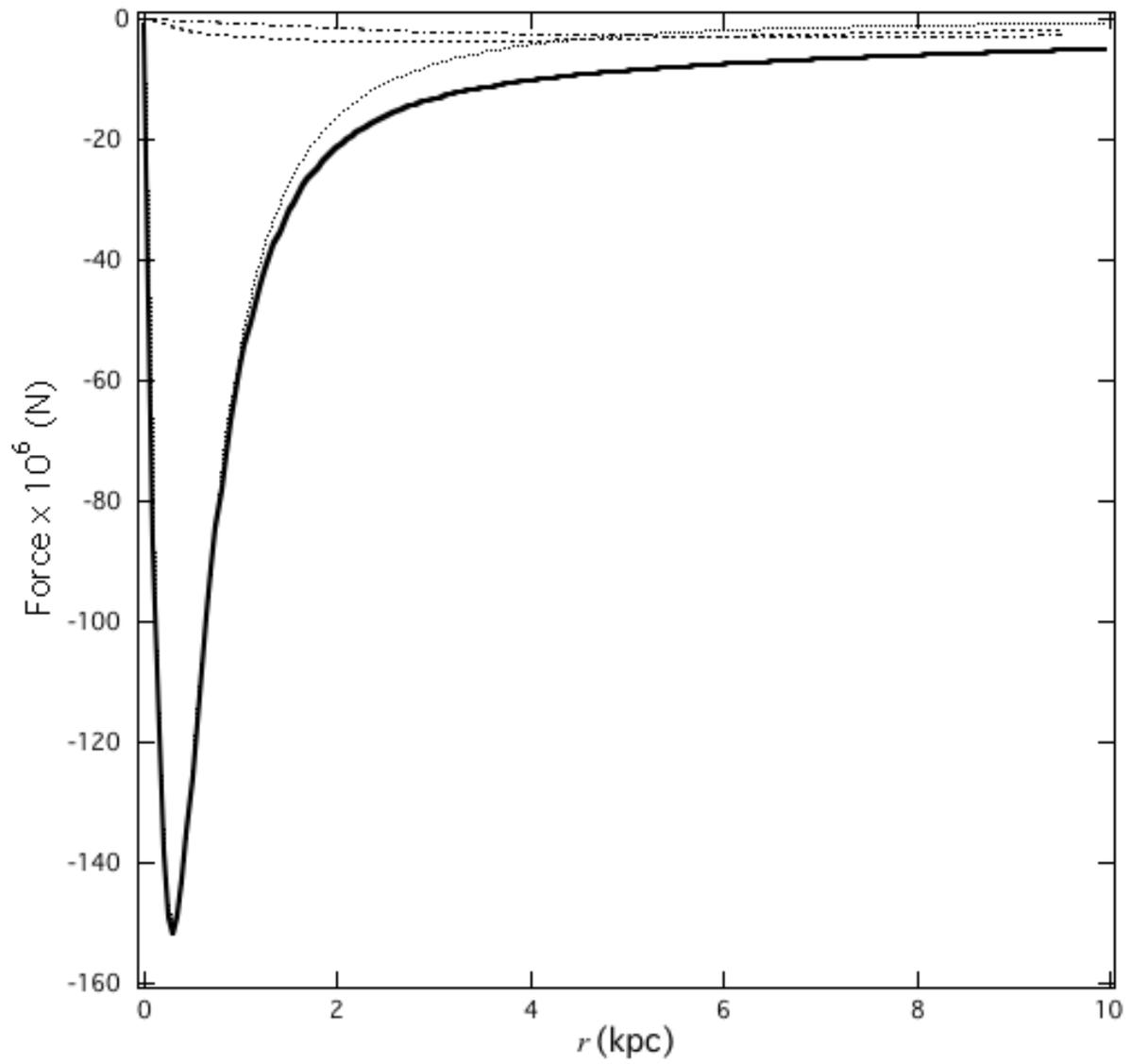

Figure 3. gravitational force vs *r*

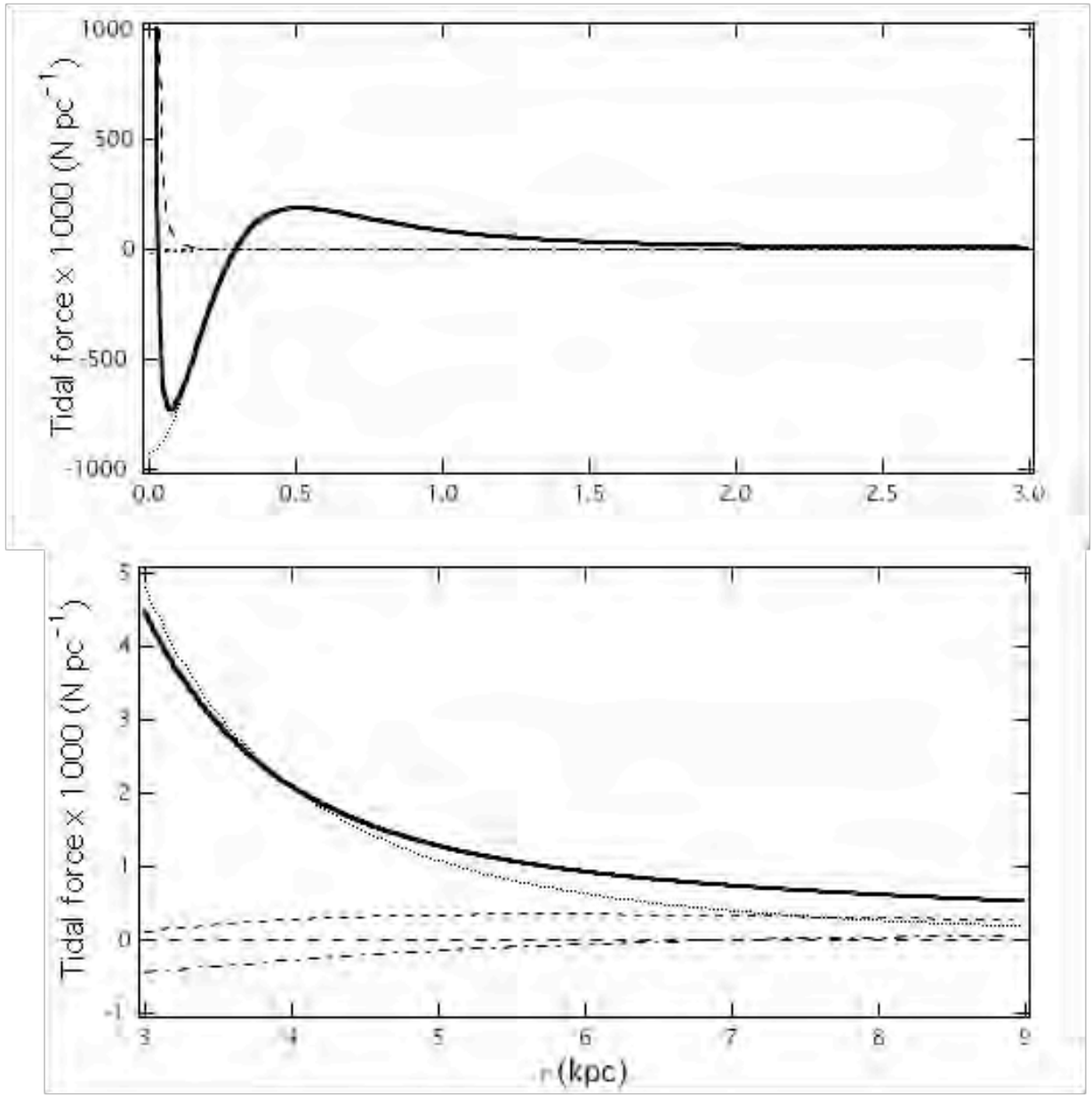

Figure 4: radial tidal force vs *r*

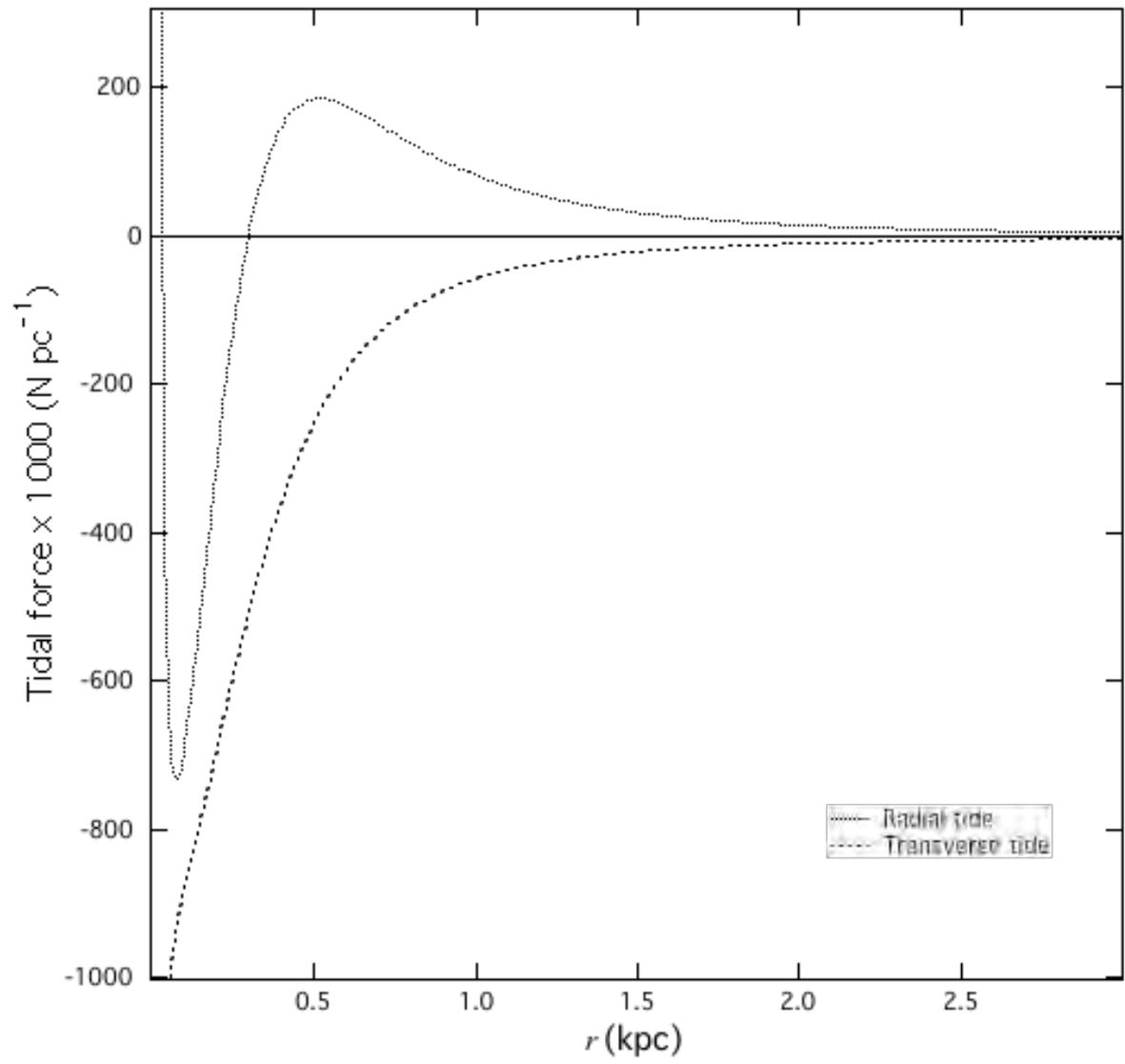

Figure 5: radial and transverse tides vs *r*

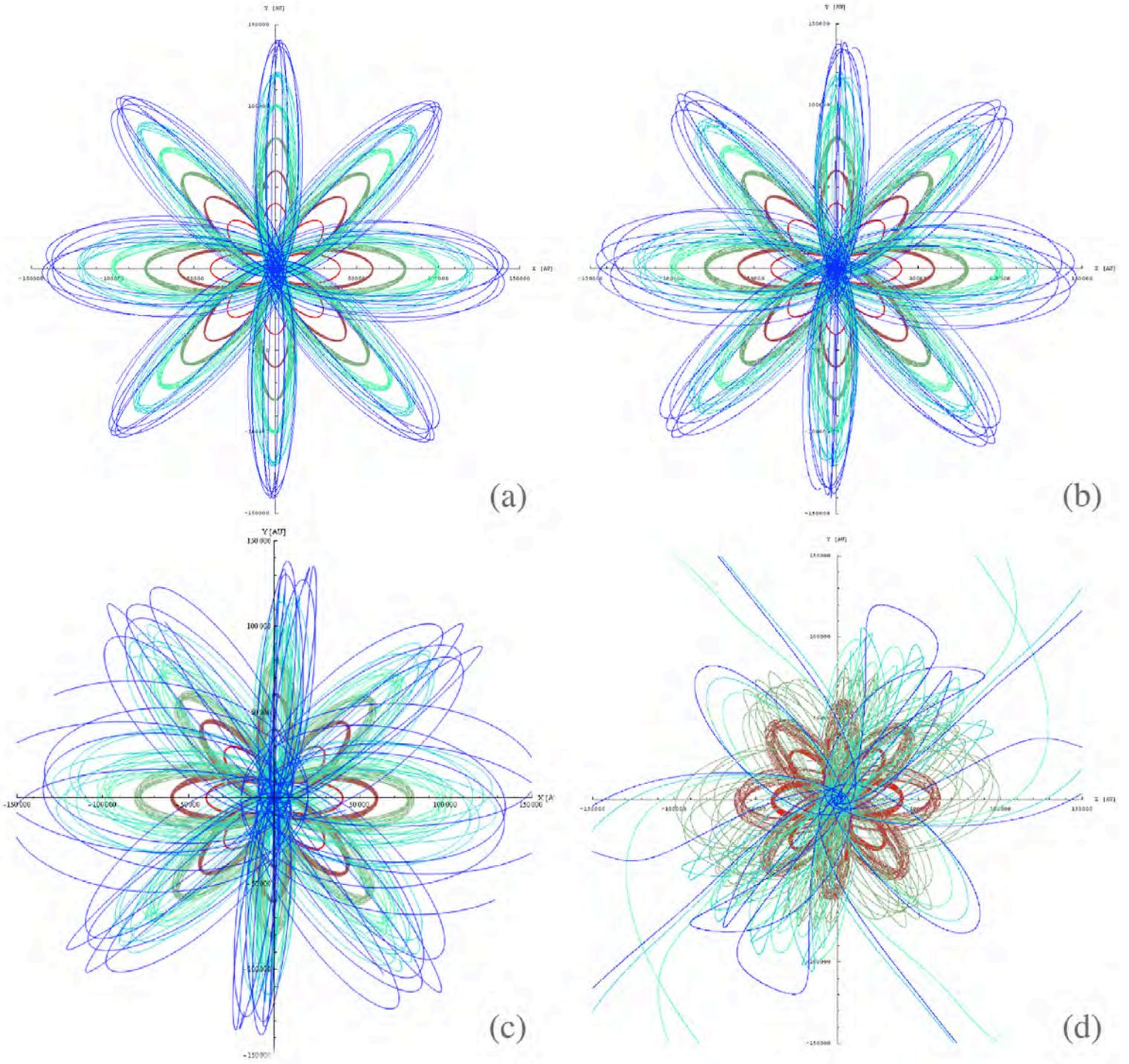

Figure 6 (a, b, c, d): 48 comets with *r* = 8, 6, 4, 2 kpc, respectively, and star mass = 1 M$_\odot$

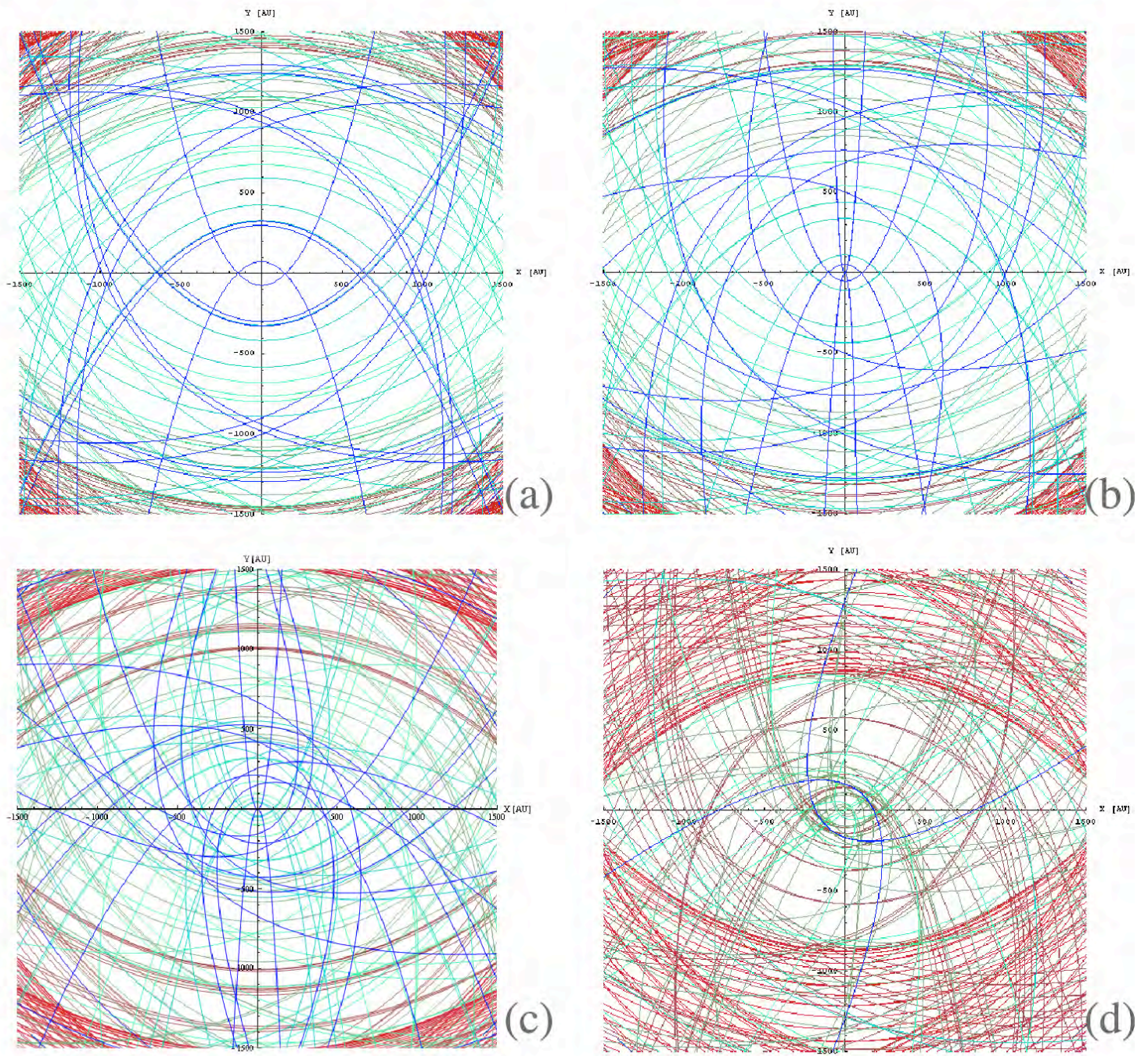

Figure 7: Same as Fig. 6; regions from -1500 to 1500 AU in x and y

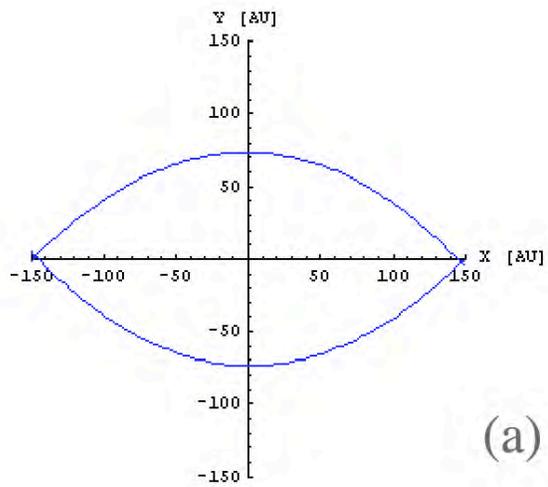
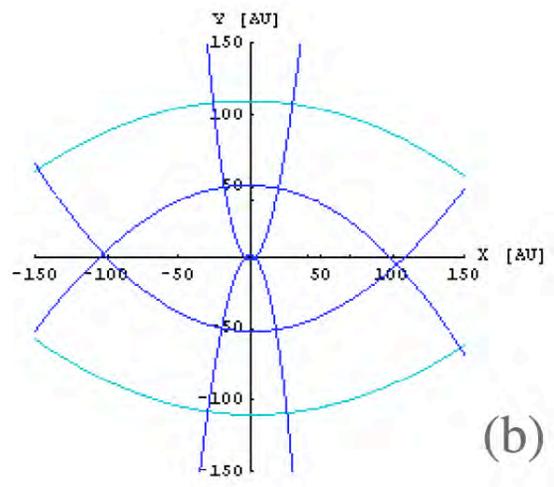
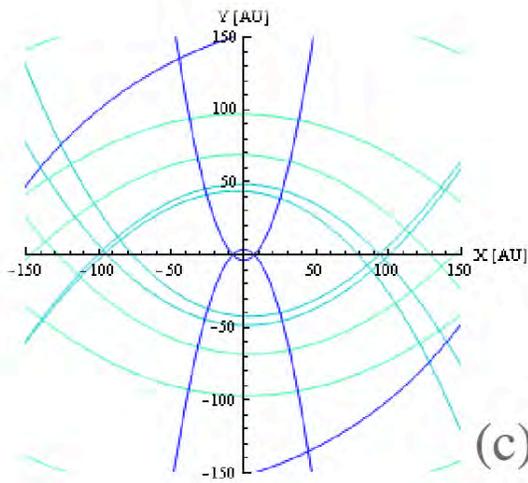
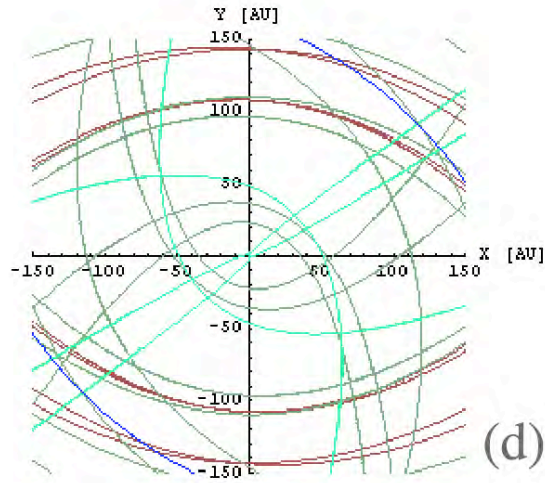

Figure 8: Same as Fig. 6; regions from -150 to 150 AU in x and y

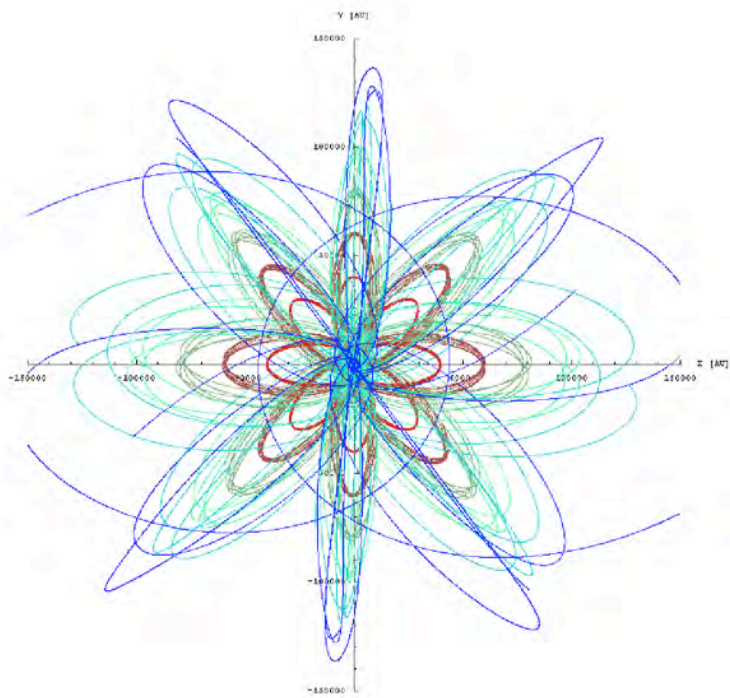

Figure 9: Comet orbits for *r* = 8 kpc and star mass 0.2 M$_\odot$

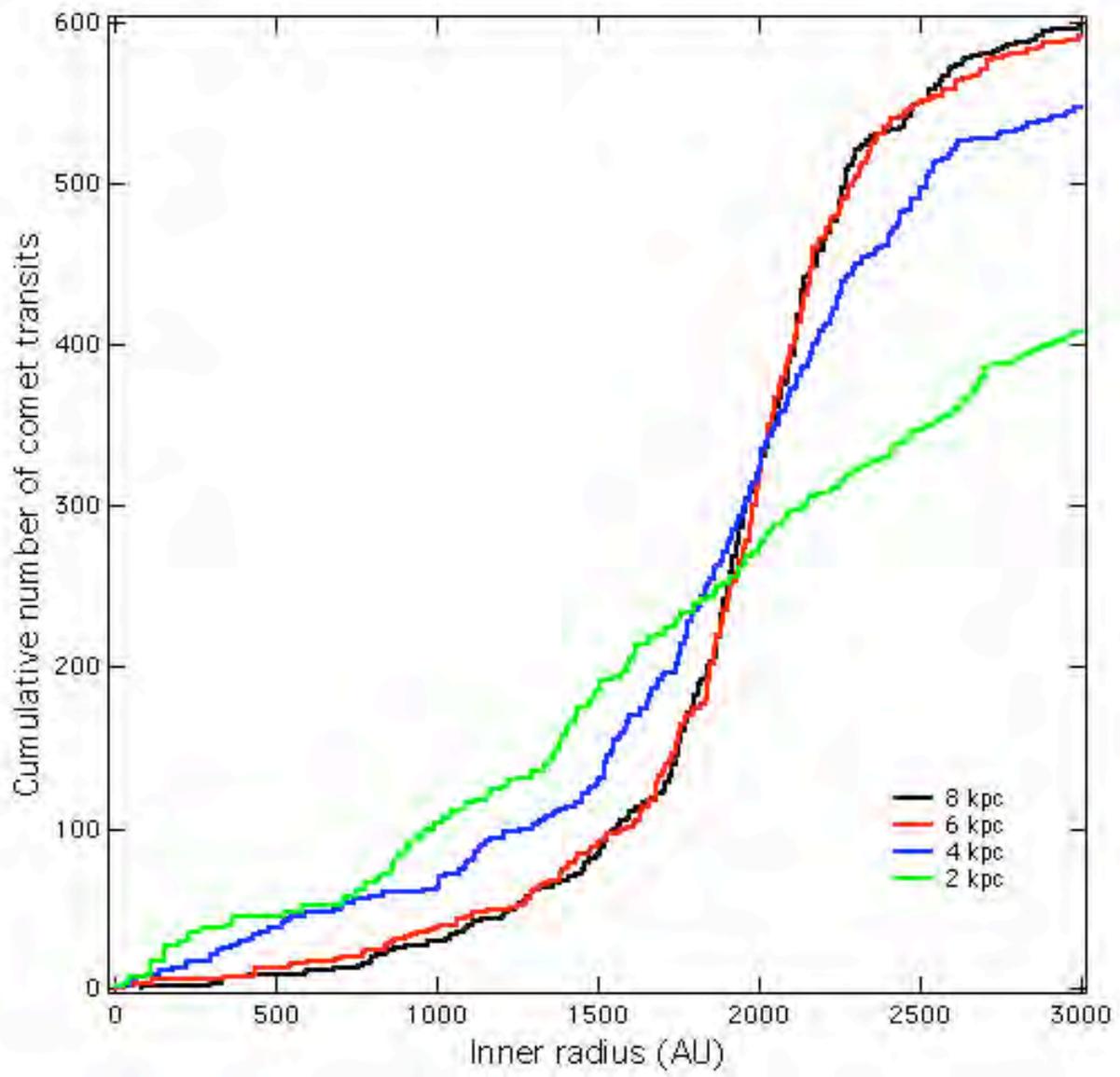

Figure 10: Cumulative count in time of the number of comet transits inside a radius of 3000 AU

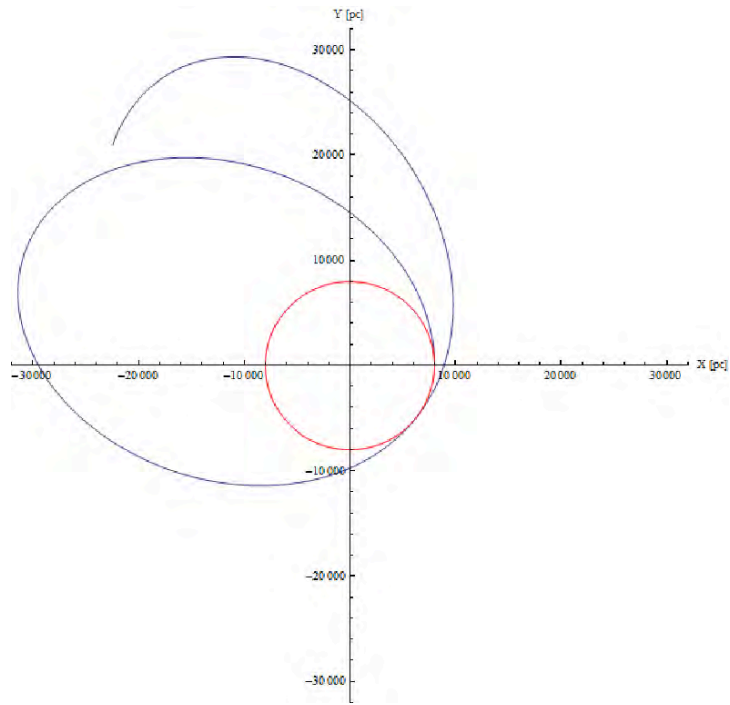

Figure 11: Sun's orbit (blue) if halo potential is switched off or left on (red).

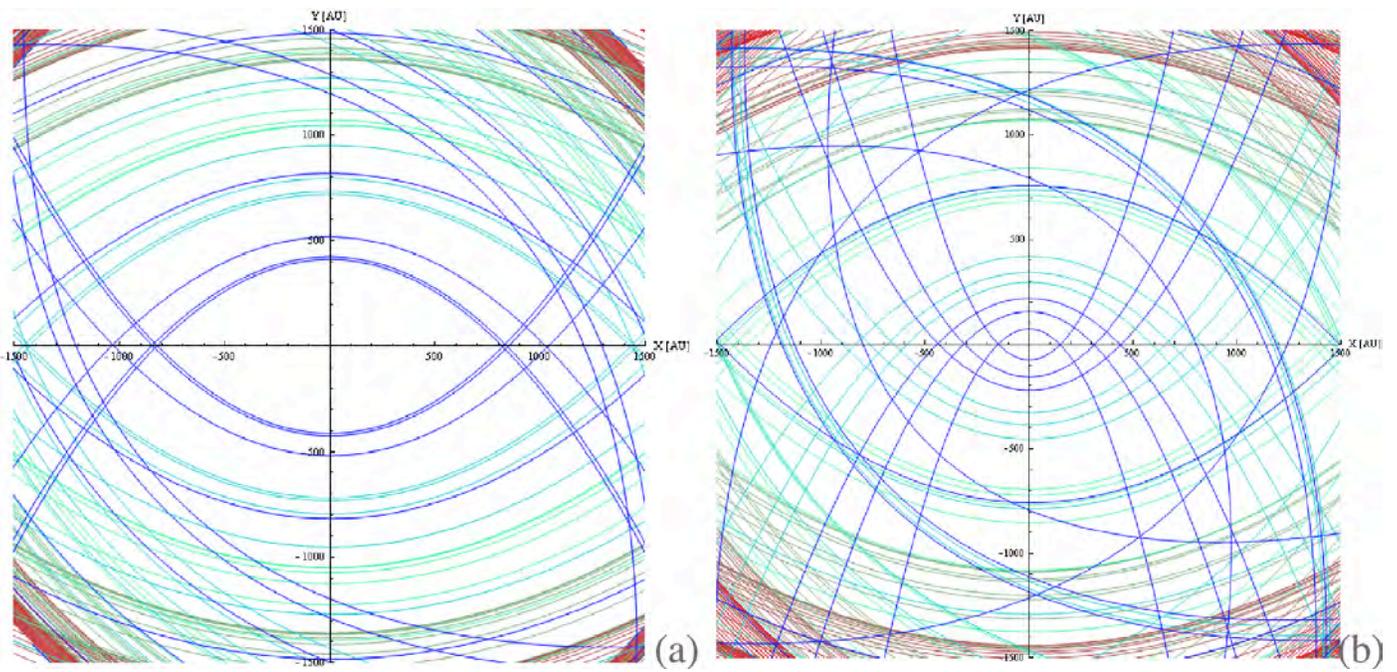

Figure 12: Inner -1500 to 1500 AUs in x and y with halo potential switched off.